\documentclass[aps,pra,preprint,superscriptaddress]{revtex4-1} 
\usepackage[latin9]{inputenc}
\usepackage{color}
\usepackage{amsmath}
\usepackage{amssymb}
\usepackage{graphicx}
\usepackage{etoolbox}
\usepackage{times}
\makeatletter

\@ifundefined{textcolor}{}
{%
 \definecolor{BLACK}{gray}{0}
 \definecolor{WHITE}{gray}{1}
 \definecolor{RED}{rgb}{1,0,0}
 \definecolor{GREEN}{rgb}{0,1,0}
 \definecolor{BLUE}{rgb}{0,0,1}
 \definecolor{CYAN}{cmyk}{1,0,0,0}
 \definecolor{MAGENTA}{cmyk}{0,1,0,0}
 \definecolor{YELLOW}{cmyk}{0,0,1,0}
}

\newcommand{\ket}[1]{{{|}{#1}\rangle}}

\begin{document}

\title{Tunable spin-spin interactions and entanglement of ions in separate wells}

\author{A. C. Wilson}
\author{Y. Colombe}
\affiliation{National Institute of Standards and Technology, 325 Broadway, Boulder, CO 80305, USA}
\author{K. R. Brown}
\affiliation{Georgia Tech Research Institute, 400 10th Street, N.W., Atlanta, GA 30332, USA}
\author{E. Knill}
\author{D. Leibfried}
\author{D. J. Wineland}
\affiliation{National Institute of Standards and Technology, 325 Broadway, Boulder, CO 80305, USA}
\maketitle

\textbf{Quantum simulation\cite{fey82,llo96}  - the use of one quantum system to simulate a less controllable one - may provide an understanding of the many quantum systems which cannot be modeled using classical computers.  Impressive progress on control and manipulation has been achieved for various quantum systems\cite{lad10,geo14,bla12}, but one of the remaining challenges is the implementation of scalable devices. In this regard, individual ions trapped in separate tunable potential wells are promising\cite{chi08,sch09,shi13}. Here we implement the basic features of this approach and demonstrate deterministic tuning of the Coulomb interaction between two ions, independently controlling their local wells. The scheme is suitable for emulating a range of spin-spin interactions, but to characterize the performance of our setup we select one that entangles the internal states of the two ions with 0.82(1) fidelity. Extension of this building-block to a 2D-network, which ion-trap micro-fabrication processes enable\cite{sei06}, may provide a new quantum simulator architecture with broad flexibility in designing and scaling the arrangement of ions and their mutual interactions. To perform useful quantum simulations, including those of intriguing condensed-matter phenomena such as the fractional quantum Hall effect, an array of tens of ions might be sufficient\cite{,fri08,nie13,geo14}.}
\noindent

The use of effective spin-spin interactions between ions in separate potential wells is a key feature of proposals for simulation with two-dimensional systems of quantum spins with arbitrary conformations and versatile couplings\cite{chi08,sch09,sch11}. In addition, these effective spin-spin interactions may enable logic operations to be performed in a multi-zone quantum information processor\cite{win98,cir00,kie02} without the need to bring the quantum bits into the same trapping potential well\cite{bro11,har11}. Such coupling might also prove useful for metrology and sensing. For example, it could extend the capabilities of quantum-logic spectroscopy\cite{hei90,sch05} to ions that cannot be trapped within the same potential as the measurement ion, such as oppositely charged ions or even antimatter particles\cite{hei90}. Coupling could be obtained either through mutually shared electrodes\cite{hei90,dan09} or directly through the Coulomb interaction\cite{win98,cia03}.
 
In the experiments described here, two ions of mass $m$ are trapped at equilibrium distance $d_0$ in independent, approximately harmonic potential wells. Coulomb interaction between the ions leads to dipole-dipole type coupling, with strength $\Omega_{\rm ex}\propto d_0^{-3}$ (see Methods), where the oscillations of the ions in their respective wells manifest the dipoles\cite{sch11}. The coupled system has six normal modes, four perpendicular to the direction between the double wells (radial) and two along this direction (axial). While all these modes are useful for dipole-dipole coupling\cite{sch11}, we concentrate on the two axial modes, with uncoupled well frequencies $\omega_l\approx \omega_r$, and with eigenfrequencies and eigenvectors
\begin{eqnarray}\label{Eq:NorMod}
\omega_{\rm str/com}&=& \bar{\omega} \pm \sqrt{\delta^2+\Omega_{\rm ex}^2}\nonumber \\
{\bf q}_{\rm str/com} &=&(\sin(\theta_{\rm str/com}),\cos(\theta_{\rm str/com})),
\end{eqnarray}
where $\theta_{\rm str/com}=\arctan[(\delta\mp \sqrt{\delta^2+\Omega_{\rm ex^2}})/\Omega_{\rm ex}]$ and the upper (lower) sign is for the str (com) mode. The average well frequency is denoted by $\bar{\omega} \equiv \frac{1}{2}(\omega_l+\omega_r)$, and the frequency difference is $2 \delta = (\omega_r-\omega_l)$. For $|\delta| \gg \Omega_{\rm ex}$ these modes decouple and the two ions move nearly independently of each other. When approaching resonance  ($\delta = 0$), the motions of the ions are strongly coupled, resulting in an avoided crossing of the motional frequencies  with a splitting of $2 \Omega_{\rm ex}$. On resonance, the normal modes are a center-of-mass mode ($\omega_{\rm com}, {\bf q}_{\rm com}=(1/\sqrt{2},1/\sqrt{2})$) and a stretch mode ($\omega_{\rm str}, {\bf q}_{\rm str}=(-1/\sqrt{2},1/\sqrt{2})$), with motional quanta shared between the two ions.

These shared quantized degrees of freedom can simulate spin-spin interactions\cite{kim10,bla12,bri12}, just as for two-qubit quantum logic gates with ions in the same harmonic well, but as opposed to the latter case, the strength of the spin-spin interaction can be tuned from strong to weak by control of the individual trapping wells\cite{sch11, bro11, har11}.  We denote the energy eigenstates of the pseudo-spin 1/2 systems as $\{\ket{\uparrow}, \ket{\downarrow}\}$, corresponding to internal states of the ions, separated by $\hbar \omega_0$, and the number states of the normal modes as $\ket{n_{{\rm str/com}}}$. We excite `carrier'-transitions $\ket{\downarrow, n_{\rm str},n_{\rm com}} \leftrightarrow \ket{\uparrow, n_{\rm str},n_{\rm com}}$ with a uniform oscillating field at the $\ket{\downarrow} \leftrightarrow \ket{\uparrow}$ transition frequency $\omega_0$, and with phase $\phi_c$. Simultaneously, a single `red sideband' excitation at frequency $\omega_0-\bar{\omega}$ and phase $\phi_s$, between the sideband frequencies for the str and com modes, excites both $\ket{\downarrow, n_{\rm str},n_{\rm com}} \leftrightarrow \ket{\uparrow, n_{\rm str} -1,n_{\rm com}}$ and $\ket{\downarrow, n_{\rm str},n_{\rm com}} \leftrightarrow \ket{\uparrow, n_{\rm str},n_{\rm com} -1}$ transitions\cite{ber12,tan13}. These excitations emulate an effective spin-spin interaction (see Methods)
\begin{eqnarray}\label{Eq:SpiHam}
\hat{H}_{\rm eff}&=& \hbar \kappa \hat{\sigma}^{\phi_c}_{l} \hat{\sigma}^{\phi_c}_{r}, \nonumber\\
\hat{\sigma}^{\phi_c}_{l/r}&=&\cos(\phi_c) \hat{\sigma}^x_{l/r}-\sin(\phi_c) \hat{\sigma}^y_{l/r},
\end{eqnarray}
where $\kappa$ is the coupling strength and $\hat{\sigma}^{x/y}_{l/r}$ are the Pauli spin-1/2 operators of the respective ions. We can emulate anti-ferromagnetic ($\kappa>0$) and ferromagnetic ($\kappa<0$) interactions by our choice of the ion spacing or the relative detunings $\delta_{\rm str}$ and $\delta_{\rm com}$ of the normal modes relative to the sideband drive (see Methods). Under the simultaneous carrier and red sideband drive, the spins become periodically entangled and disentangled with the motion. Starting with a product state $\ket{\Phi_i}$, spins and motion are disentangled into a product state at $T_j= 2\pi ~j/\Omega_{\rm ex}$ ($j>0$ integer), but the spins acquire phases that depend on the ions' motion in phase space during the off-resonant excitation.  These phases simulate the spin-spin interaction \cite{por04}. We benchmark our implementation of the spin-spin interaction by starting from the well-defined product state $\ket{\Psi_i} =\ket{\downarrow \downarrow}$, effectively evolving it under an anti-ferromagnetic ($\kappa>0$) interaction for time $ T_2 =\pi/(4 \kappa)$, $\phi_c=0$ and comparing the resulting state to the maximally entangled state  $\ket{\Psi_e}=\exp[-i \frac{\pi}{4}\hat{\sigma}^x_{l}\hat{\sigma}^x_{r} ]\ket{\Psi_i}=\frac{1}{\sqrt{2}} (\ket{\downarrow \downarrow} - i \ket{\uparrow \uparrow})$ that would be produced under ideal conditions (see Methods).

\begin{figure}
\includegraphics[width=10 cm]{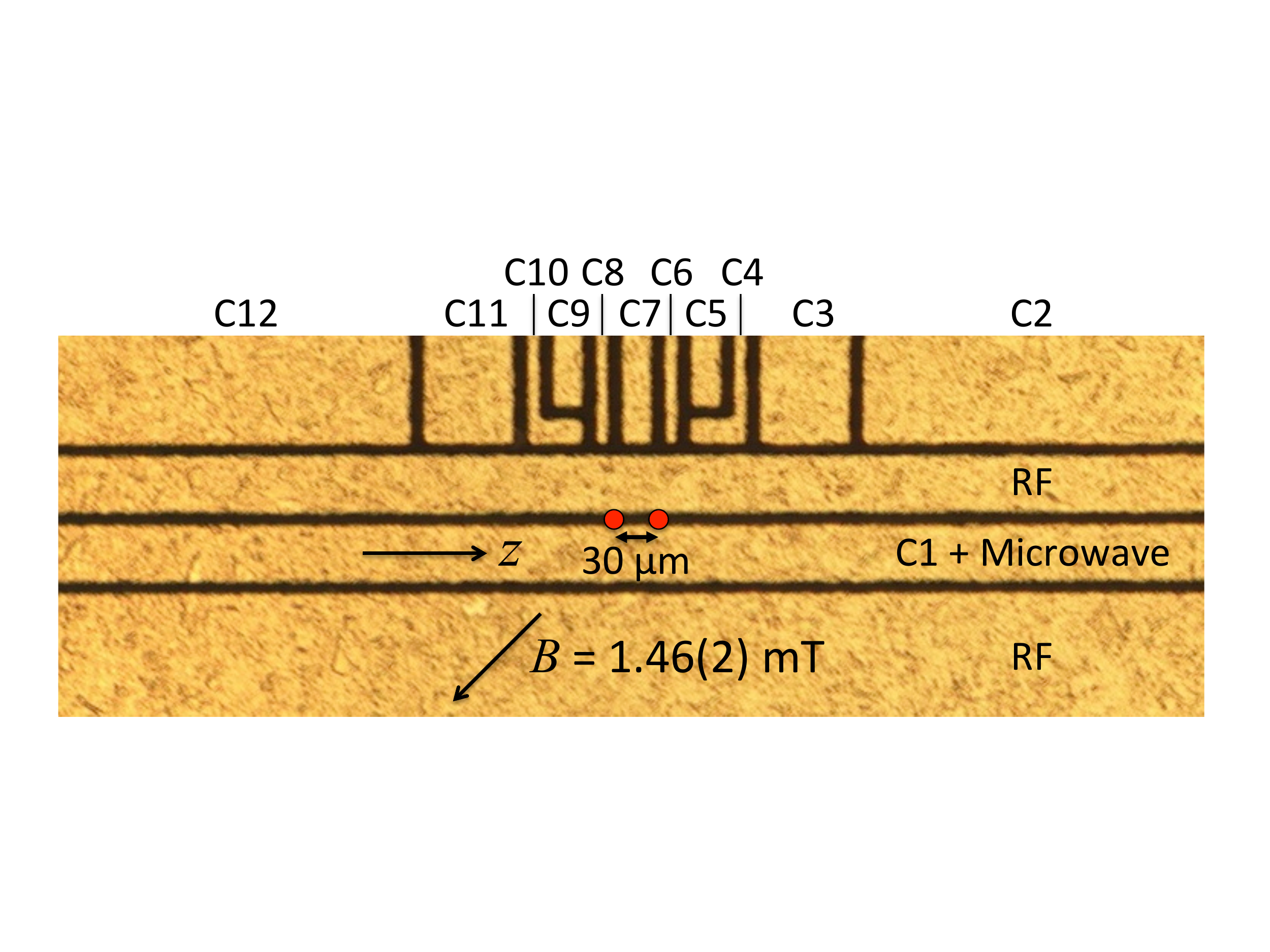}
\caption{\textbf{Micro-fabricated surface-electrode trap.} Microscope image of ion trap electrodes, showing radio-frequency (RF) and static-potential control electrodes (C1$\--$C12).  Dark areas are the 5 $\mu$m gaps between electrodes.  Ions are trapped 40 $\mu$m above the chip surface; red dots indicate the ion locations with a 30 $\mu$m spacing. Electrode C1 also supports microwave currents at 1.28 GHz to drive carrier transitions on the two ions.}
\label{fig:Trap}
\end{figure}

 The (pseudo-) spin 1/2-system is formed by the $\ket{2s~^2S_{1/2}, F=1,m_F=-1} \equiv \ket{\uparrow}$ and $\ket{2s~^2S_{1/2}, F=2,m_F=-2} \equiv \ket{\downarrow}$ hyperfine ground states of $^9$Be$^{+}$, where $F$ is the total angular momentum and $m_F$ is the component of $F$ along a quantization axis provided by a 1.46(2) mT static magnetic field (see Fig.~\ref{fig:Trap}). The ions are confined in a cryogenic (trap temperature $<$ 5 K), micro-fabricated, surface-electrode linear Paul ion trap~\cite{bro11} composed of 10 $\mu$m thick gold electrodes separated by 5  $\mu$m gaps, deposited onto a crystalline quartz substrate.  An oscillating potential ($\sim$100 V peak at 163 $\mathrm{MHz}$), applied to the RF electrodes in Fig.~\ref{fig:Trap}, provides pseudo-potential confinement of the ions in the radial (perpendicular to $z$) directions at motional frequencies of $\sim$$17\,\mathrm{MHz}$ and $\sim$$27\,\mathrm{MHz}$ at a distance of approximately 40 $\mu$m from the trap surface. Along the trap $z$-axis, a double well is formed by static potentials applied to control electrodes C1 through C12.  The axial ($z$) oscillation frequencies $\omega_l, \omega_r$ around the respective minima are typically near $4\,\mathrm{MHz}$. Single-ion heating \cite{win98} is in the range of 100 to 200 quanta per second. This heating is approximately four orders of magnitude larger than that due to our estimate of Johnson noise heating for this apparatus.  For two ions spaced 30 $\mu$m apart, and in motional resonance ($\delta = 0$), the period required for the ions to exchange their motional energies is $\tau_{\rm ex} \equiv \pi/2\Omega_{\rm ex}$ = 70 $\mathrm{\mu s}$, compared with 5 to 10 $\mathrm{ms}$ average period to absorb a single motional quantum due to background heating. Fine adjustment of control-electrode potentials (at the 100 $\mu$V level) enables individual control of potential-well curvatures to tune the Coulomb interaction between the ions through resonance. Electrode C1 also supports microwave currents (typically of milliampere amplitude) that produce an oscillating magnetic field to drive carrier transitions at the same rate in both ions.

Superimposed $\sigma^-$-polarized laser beams, nearly resonant with  the $2s~^2S_{1/2} \rightarrow 2p~^2P_{1/2}$ and the $2s~^2S_{1/2} \rightarrow 2p~^2P_{3/2}$ transitions ($\lambda \simeq$ 313 nm) and propagating along the B-field direction, are used for optical pumping, Doppler laser cooling, and state detection by resonance fluorescence. Optical pumping prepares both ions in $\ket{\downarrow}$. We can distinguish the $\ket{\downarrow}$ (bright) and $\ket{\uparrow}$ (dark) state by detecting resonance fluorescence on the $\ket{\downarrow}\rightarrow \ket{2p~^2P_{3/2},F=3,m_F=-3}$ optical cycling transition. Typically, 3 to 5 photons are detected per ion in $\ket{\downarrow}$ over a background of 0.15 to 0.6 photons on a photomultiplier during detection periods in the range 300 to 400 $\mu$s. A pair of elliptically shaped laser beams, separated in frequency by approximately the $\ket{\downarrow} \leftrightarrow \ket{\uparrow}$ transition frequency ($\omega_0\simeq 2\pi \times$ 1.28 GHz) and detuned 80 GHz above the $^2S_{1/2}~\rightarrow~^2P_{1/2}$ transition, illuminate both ions with equal intensity.  These beams induce two-photon stimulated-Raman transitions for ground-state cooling\cite{win98} and for the motional sideband excitations used to implement the spin-spin interaction \cite{por04}. Derived from the same 313 nm source, the frequency difference between the beams is produced with acousto-optic modulators, and the beam orientation is such that the difference wave-vector $\bf{k}={\bf k}_2-{\bf k}_1$ is parallel to the $z$-axis (with magnitude $k = 2 \sqrt{2} \pi/\lambda$).  The spin-spin coupling strength is $\kappa=\cos( 2\phi)(\eta \Omega_s)^2/(2 \Omega_{\rm ex})$, where $2\phi = kd_0$ is the phase difference of the beat-note between the two laser fields at the positions of the ions, $\Omega_s$ is the stimulated-Raman Rabi frequency, and $\eta=k/\sqrt{\hbar/(2 m \bar{\omega})}$ (see Methods).


\begin{figure}
\includegraphics[width=10 cm]{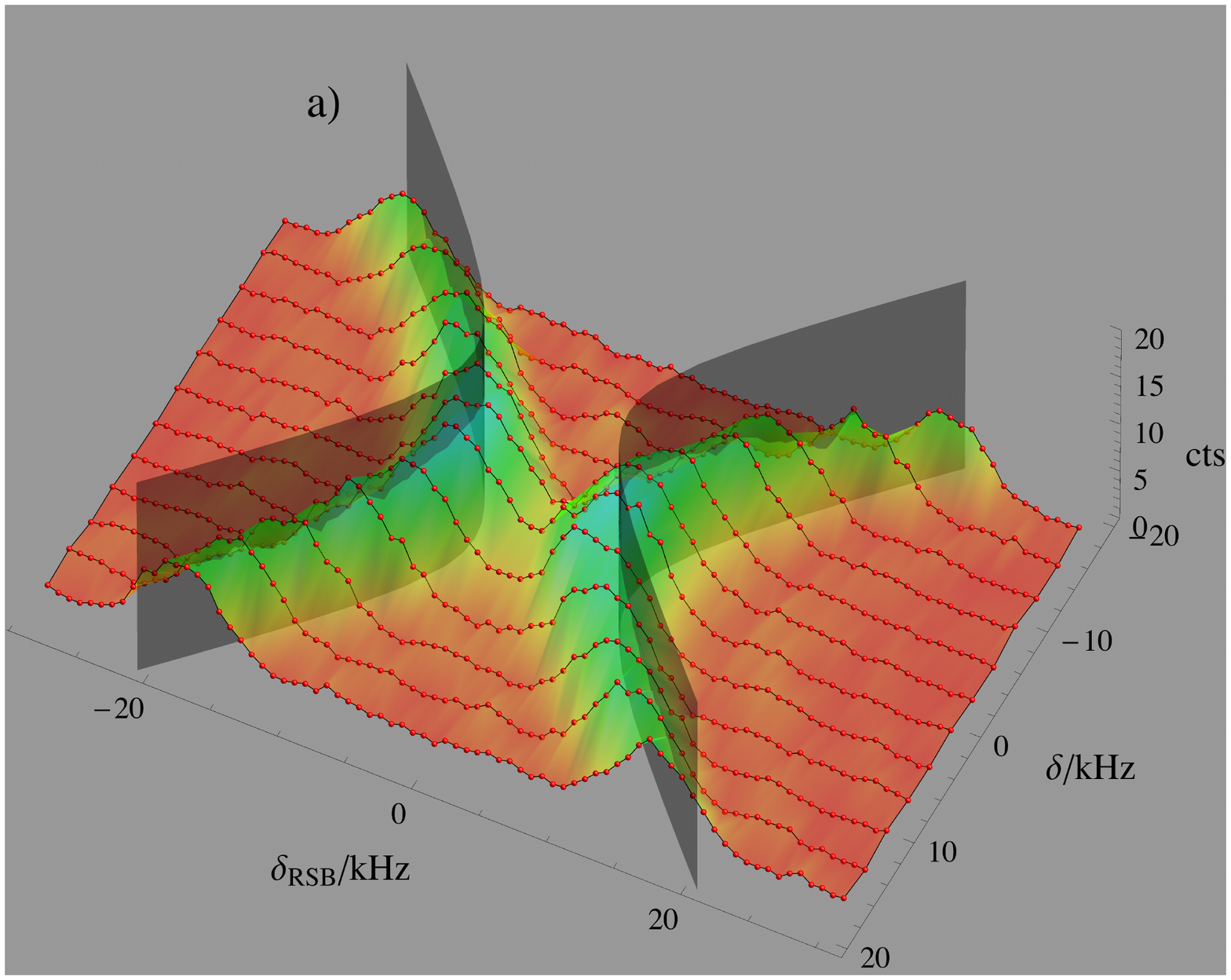}
\includegraphics[width=10 cm]{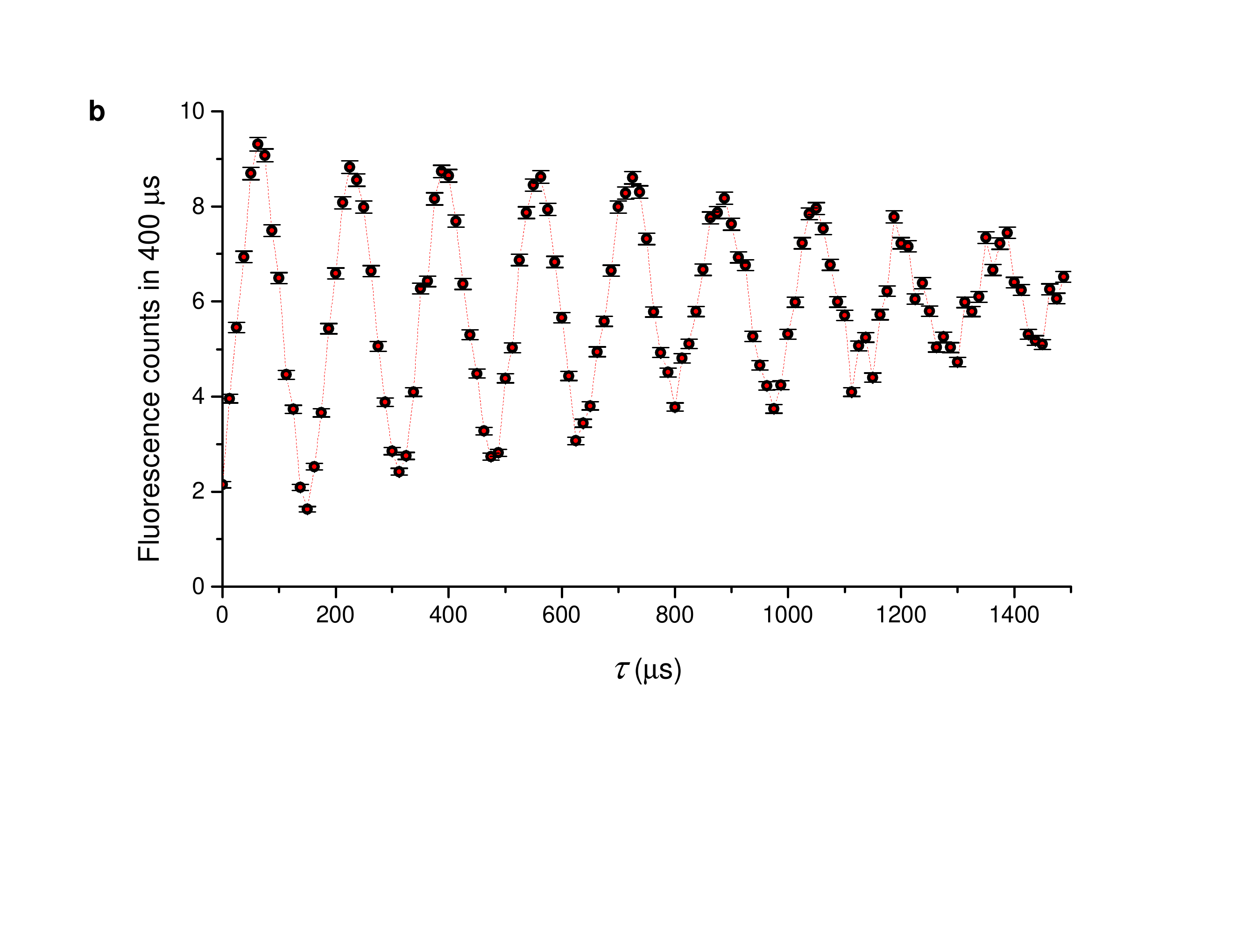}
\caption{\textbf{Motional spectroscopy of two coupled ions.} {\bf a}, The red dots connected by black lines indicate separate scans of the red sideband detuning $\delta_{\rm RSB}$ from the average mode frequency $\bar{\omega}$ for different values of the difference $\delta$ between the individual well frequencies.  The vertical scale is proportional to the sum of the probabilities for each ion to be in $\ket{\downarrow}$. At the center of the avoided crossing the normal mode frequency splitting $ \Omega_{\rm ex}/\pi$ is 12(1) kHz.  Each data point represents an average of 200 experiments. Shaded planes are a theoretical prediction for the avoided crossing according to Eqs.(\ref{Eq:NorMod}). {\bf b}, Resonant ($\delta\approx 0$) single-quantum motional exchange between two ions, with an exchange time $\tau_{ex}$  = 80(2) $\mu$s. The vertical scale is proportional to the probability of the laser-addressed ion being in $\ket{\downarrow}$.  Each data point represents an average of 500 experiments, and error bars correspond to standard error of the mean.  Dashed lines are included to guide the eye.}
\label{fig:CroExc}
\end{figure}

A key to implementing spin-spin interactions with ions in separate trapping zones is being able to tune the well frequencies precisely enough to control the eigenfrequencies and eigenmodes (Eqs.~(\ref{Eq:NorMod})) near the avoided crossing.  In Fig.~\ref{fig:CroExc}a, we characterize this avoided crossing. For these experiments, the ions are separated by 27(2) $\mu$m. They are laser cooled nearly to their motional ground states (mean motional mode occupation $\bar{n}_{\rm str/com} \approx 0.1$), optically pumped to the $\ket{\downarrow \downarrow}$ state and then rotated into the $\ket{\uparrow \uparrow}$ state with a microwave carrier $\pi$-pulse. Fine adjustments are made to control electrodes C2 and C12 to tune the harmonic confinement of the two trapping zones, stepping the system through the avoided crossing.  At each step, after cooling and optical pumping, we implement the Raman red-sideband drive and scan its detuning $\delta_{\rm RSB}$ with respect to $\bar{\omega}$.  If the sideband excitation frequency is equal to $\omega_0 - \omega_{\rm str}$ or $\omega_0 - \omega_{\rm com}$, the spin of one or both ions can flip to $\ket{\downarrow}$ while absorbing quanta of motion, and a peak in the resonance fluorescence counts is observed. The spectral resolution is set by the duration of the square-pulse sideband excitation (120 $\mu$s).  At the center of the avoided crossing the splitting of the mode frequencies is $2 \Omega_{\rm ex} = 2 \pi \times 12(1)$ kHz.

In Fig.~\ref{fig:CroExc}b we show data that demonstrate single-phonon exchange between the two ions. With the trapping zones tuned to resonance ($\delta = 0$), both modes are cooled to near the motional ground state and the ions prepared in $\ket{\uparrow \uparrow}$. In this experiment, the two Raman beams are tightly focussed onto only one of the ions and used to add a single phonon to that ion (and flip its spin) with a $\pi$-pulse on the red sideband of its local frequency in a duration short compared to $\tau_{\rm ex}$.  In this limit, after the pulse, the resulting motional state is an equal superposition of both modes, therefore the phonon energy is exchanged back and forth between the ions with a period 2$\tau_{\rm ex}$ \cite{bro11}. To monitor the exchange, the same Raman interaction is applied again after a variable delay $\tau$. This can  flip the spin and remove the quantum of motion only if the motion resides solely in the addressed ion after a particular delay. The level of fluorescence is proportional to the probability of this spin flip. From this, we determine an exchange time $\tau_{\rm ex}$  = 80(2) $\mu$s, consistent with an ion spacing of 30(2) $\mu$m for this experiment.  The reduction in contrast for longer delays is caused mainly by fluctuations and drifts of the trapping potential. We estimate that $\delta/(2 \pi)$ drifted by  approximately 500 Hz (a significant fraction of $\Omega_{\rm ex}/(2 \pi)~$) during the 2 to 3 minutes required for the 20,000 experiments that provide the data for Fig.~\ref{fig:CroExc}b.


\begin{figure}
\includegraphics[width=12 cm]{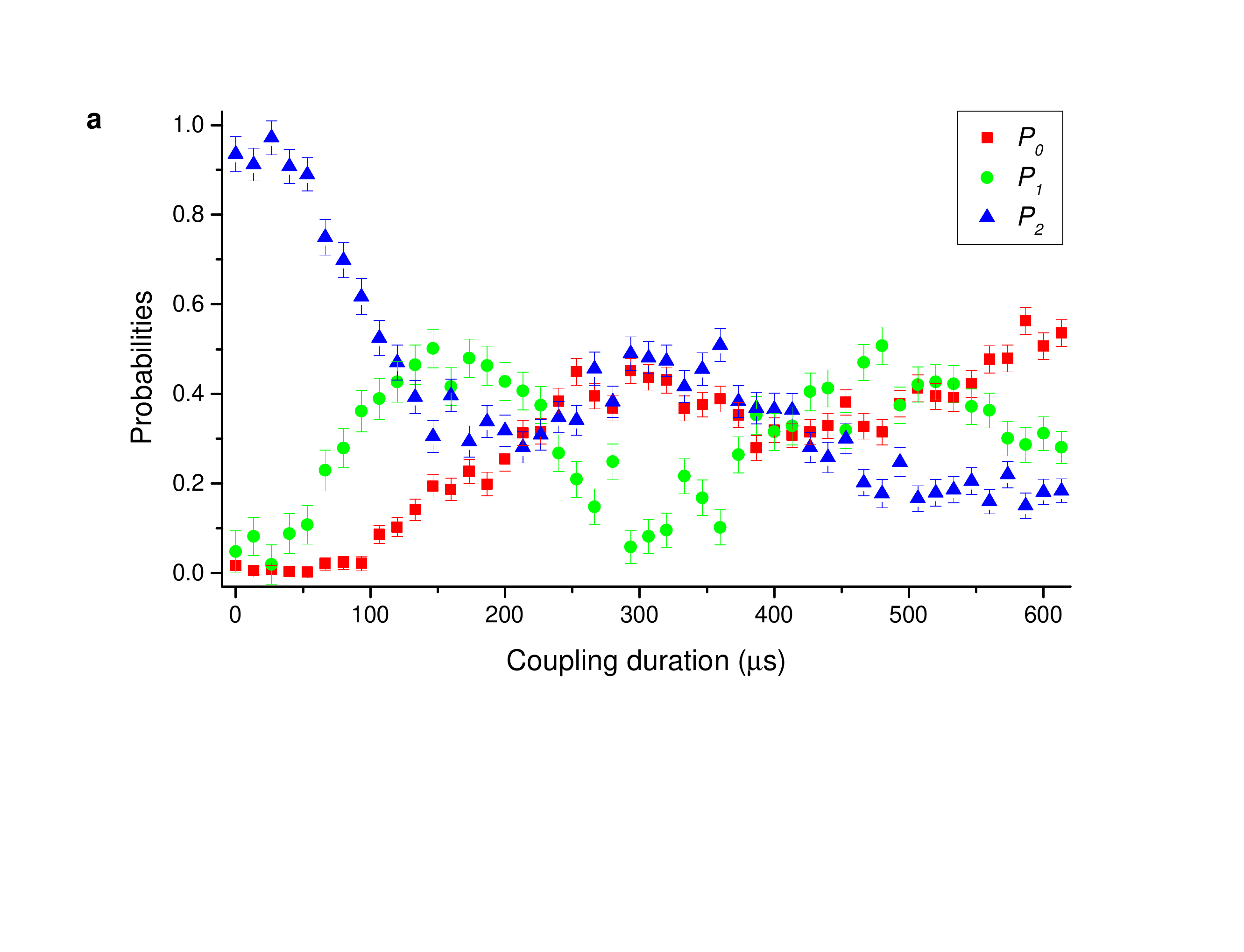}
\includegraphics[width=10 cm]{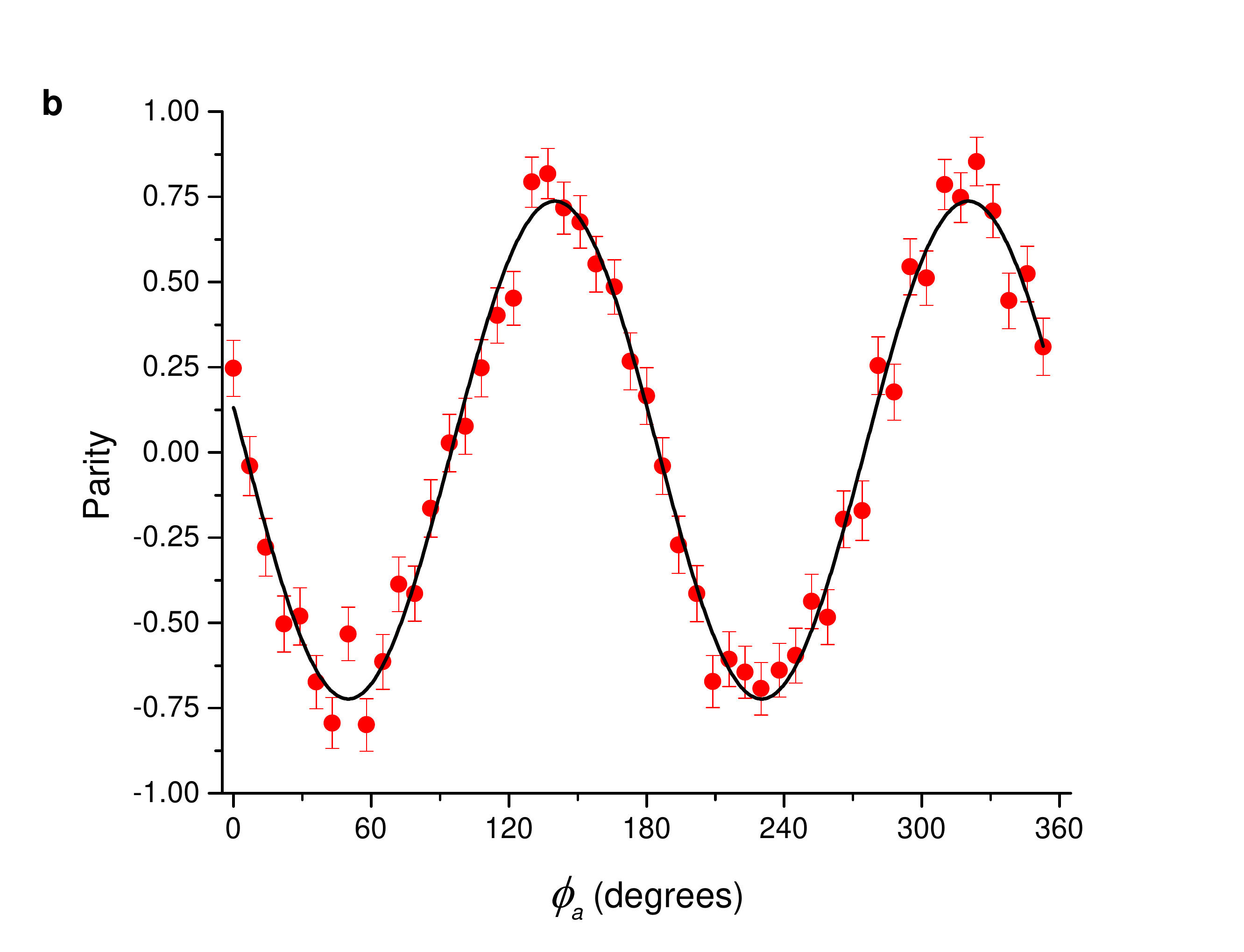}
\caption{\textbf{Characterizing the spin-spin coupling interaction between ions in separate trapping zones.} {\bf a}, Evolution of probabilities $P_0$ of $\ket{\uparrow \uparrow}$ (red), $P_1$ of $\ket{\downarrow \uparrow}$ and $\ket{\uparrow \downarrow}$ (green), and $P_2$ of $\ket{\downarrow \downarrow}$ (blue), as a function of coupling duration. Each data point represents an average of 400 experiments, and error bars correspond to standard error of the mean. {\bf b}, Parity oscillation obtained (for a coupling duration of 300 $\mu s$) by applying a $\pi$/2-carrier analysis pulse with variable phase $\phi_a$, and a fit to the data (black curve).  Each data point represents an average of 400 experiments, and error bars correspond to standard error of the mean.}
\label{fig:PopPar}
\end{figure}

For benchmarking the spin-spin interaction, the laser beams for fluorescence detection, Doppler cooling and stimulated Raman transitions are made to spatially overlap both ions with equal intensity. The ion spacing (approximately 27 $\mu$m here) is adjusted to an integer number of half-wavelengths of the difference wave-vector of the two Raman laser fields, by a technique described elsewhere\cite{cho11}, such that $\cos(2\phi) \approx 1$.  The wells are tuned on resonance  ($\delta = 0$) with adjustments to control electrodes C2 and C12. The ions are first Doppler cooled, then Raman sideband cooled to near the ground state on both normal modes, and finally optically pumped into the $\ket{\downarrow \downarrow}$ state. The spin-spin interaction is implemented by simultaneously applying a relatively strong resonant microwave carrier excitation (Rabi frequency, $\Omega_c$ = 2$\pi \times$ 23.1(2) kHz) and an optical sideband excitation at $\omega_0-\bar{\omega}$ (Rabi frequency, $\eta \Omega_s$ = 2$\pi \times$ 2.4(2) kHz).  The exchange frequency, $2 \Omega_{\rm ex} = 2 \pi \times 13(1)$ kHz, so that $\kappa = 2 \pi \times 446(13)$ Hz. In the middle of the coupling period, we shift the phases $\phi_c$ and $\phi_s$ of both driving fields by 180$^\circ$ relative to the first half. These phase reversals suppress the dependence of the final state on the carrier Rabi frequency and reduce sensitivity of the spin-spin interaction to drifts in the detuning and the coupling time (see Methods). At the end of the coupling period, fluorescence detection and subsequent fitting of the photon-count histograms to those for the three possible outcomes (two ions bright, $\ket{\downarrow \downarrow}$;  one ion bright, $\ket{\downarrow \uparrow}$ or $\ket{\uparrow \downarrow}$; or both ions dark, $\ket{\uparrow \uparrow}$) yield the respective probabilities $P_2, P_1$ and $P_0$.

Evolution of these probabilities as a function of the coupling duration is shown in Fig.~\ref{fig:PopPar}a. Near 300 $\mu$s, $P_2$ and $P_0$ are approximately equal ($P_2+P_0 = 0.91(2)$) and $P_1$ has reached a minimum. To show that the resulting state is entangled, in a subsequent experiment we stop the evolution at 300 $\mu$s, apply a carrier $\pi/2$-pulse of variable phase $\phi_a$,  and determine the parity $\Pi=P_2+P_0-P_1$ as a function of $\phi_a$. These data are shown in Fig.~\ref{fig:PopPar}b together with a fit to  $A\cos(2\phi_a+\phi_0)+B$.  The fitted probabilities and the contrast $A=0.73(2)$ imply a state fidelity \cite{sac00} $F=\langle\Psi_e|\rho_e|\Psi_e\rangle=\frac{1}{2}(P_2+P_0+A) = 0.82(1)$, where the density matrix $\rho_e$ describes the experimentally produced state (see Methods). From simulations and independent measurements, we estimate the leading contributions to the observed infidelity as follows: drift and fluctuations of the trapping potentials (including ``anomalous'' motional heating) $\approx 0.08$; spontaneous emission due to off-resonant excitation by Raman laser beams $\approx 0.02$; Raman laser beam intensity fluctuations  $\approx 0.03$; and state preparation and detection errors $\approx 0.03$.

For scalable implementations of lattices of interacting spins, the quality and ease of tuning of the spin-spin interaction must be improved; however, there are no apparent fundamental barriers. Trap potential fluctuations in our experiments appear to be dominated by changes in surface charging and work functions rather than changes in externally applied control-potentials.  It should be possible to suppress these fluctuations by improving the surface quality of the electrodes \cite{hit12}, reducing the amount of nearby dielectric materials, and minimizing the exposure of the electrodes to UV light through better beam shaping.  Laser intensity and pointing noise can be reduced by passive and/or active stabilization of the beams with respect to the ions, or potentially avoided entirely by utilizing microwave gradient fields for the sideband interactions\cite{sch11}.  The micro-fabrication techniques used to construct the trap are scalable to larger arrays of trapped ions, thus potentially enabling informative ``analog'' quantum simulations\cite{geo14} without requiring arbitrarily precise quantum control. Theoretical work to quantify the common belief that many observables of interest in analog quantum simulations are sufficiently robust is ongoing\cite{hau12}. Initial indications are that the proposed technical improvements may well be sufficient.  A 3-by-3 lattice is sufficient to simulate quantum Hall physics, and with 6-by-6 lattices fractional Hall effects and other intriguing solid state phenomena become accessible\cite{shi13,nie13}.  Even for these modest numbers of spins, modeling of quantum interactions with conventional computers is challenging; this difficulty may be overcome with quantum simulations.

\newpage
\section*{Acknowledgments}
We thank Katharine McCormick, Adam Keith and David Allcock for comments on the manuscript. This research was funded by the Office of the Director of National Intelligence (ODNI), Intelligence Advanced Research Projects Activity (IARPA), ONR, and the NIST Quantum Information Program. All statements of fact, opinion or conclusions contained herein are those of the authors and should not be construed as representing the official views or policies of IARPA or the ODNI.  This work, a submission of NIST, is not subject to US copyright.

\section*{Author Contributions}
ACW and DL designed the experiment, developed components of the experimental apparatus, collected data, analyzed results and wrote the manuscript.  DL developed the theory.  YC fabricated the ion trap chip. KRB built components of the apparatus, most notably the cryostat, and participated in the early design phase of the experiment. EK assisted with data analysis. DJW participated in the design and analysis of the experiment. All authors discussed the results and the text of the manuscript.

\section*{Author Information}
Reprints and permissions information are available at www.nature.com/reprints. The authors declare no competing financial interests. Readers are welcome to comment on the online version of this article at www.nature.com/nature. Correspondence and requests for materials should be addressed to ACW (andrew.wilson@nist.gov).


\newpage

\section*{METHODS}

\subsection*{Normal modes of the coupled wells}
We consider two ions, cooled close to their motional ground states. Along the direction of separation, each ion is confined to a separate minimum of a double-well potential with minima denoted $l$ (left) and $r$ (right). We assume much stronger confinement in the remaining directions, such that it is sufficient to consider only motion along the direction of the separated double well. The Hamiltonian of the motion of two ions of mass $m$ and charge $Q$, spaced at an average distance $d_0$ in wells with local harmonic oscillator ladder operators $\hat{a}_l$ and $\hat{a}_r$ and uncoupled oscillation frequencies $\omega_l$ and $\omega_r$, including Coulomb coupling and neglecting constant energy terms, can be written for small motional excitation \cite{bro11}
\begin{equation}\label{AEq:CouInt}
\hat{H}_{\rm m} = \hbar \omega_l \hat{a}_l^\dag \hat{a}_l+  \hbar \omega_r \hat{a}_r^\dag \hat{a}_r-\hbar \Omega_{\rm ex}(\hat{a}_l^\dag \hat{a}_r+ \hat{a}_r^\dag \hat{a}_l),
\end{equation}
with
\begin{equation}\label{AEq:CouRab}
\Omega_{\rm ex}= \frac{Q^2}{4 \pi \epsilon_0 m \sqrt{\omega_l \omega_r} d_0^3}.
\end{equation}
We define $\bar{\omega} \equiv \frac{1}{2}(\omega_l+\omega_r)$ and $\delta \equiv \frac{1}{2}(\omega_r-\omega_l)$ and transform the motion into a normal-mode basis with eigenfrequencies and eigenvectors (expressed in the eigenmode basis of two uncoupled ions)
\begin{eqnarray}\label{AEq:NorMod}
\omega_{\rm str/com} &=& \bar{\omega} \pm \sqrt{\delta^2+\Omega_{\rm ex}^2},\nonumber\\
{\bf q}_{\rm str/com}& =& (q^{(l)}_{\rm str/com}, q^{(r)}_{\rm str/com}) = (\sin (\theta_{\rm str/com}), \cos (\theta_{\rm str/com})), 
\end{eqnarray}
where
\begin{equation}
\theta_{\rm str/com} = \arctan\left[\frac{\delta \mp \sqrt{\delta^2+\Omega_{\rm ex}^2}}{\Omega_{\rm ex}}\right],
\end{equation}
and the upper (lower) sign applies to the str (com) mode. In this basis the motional Hamiltonian is
\begin{equation}
\hat{H}_{\rm m} = \hbar \omega_{\rm str} \hat{a}_{\rm str}^\dag \hat{a}_{\rm str} +  \hbar \omega_{\rm com} \hat{a}_{\rm com}^\dag \hat{a}_{\rm com},
\end{equation}
where $\hat{a}_{\rm str/com}$ are the corresponding ladder operators in the coupled basis. For $\delta=0$ we recover the familiar center-of-mass (COM) and stretch modes with a mode splitting of $2 \Omega_{\rm ex}$, and in the limit $\delta \ll \Omega_{\rm ex}$ we can approximate
\begin{equation}\label{AEq:NorApp}
{\bf q}_{\rm str/com} \approx \left( \frac{\mp 1}{\sqrt{2}}\left(1 \mp \frac{\delta}{2 \Omega_{\rm ex}} \right), 
 \frac{1}{\sqrt{2}}\left(1 \pm \frac{\delta}{2 \Omega_{\rm ex}} \right)\right).
\end{equation}
\subsection*{Interaction Hamiltonian including internal states and a combined carrier/sideband drive}
The two ions are driven resonantly by a spatially uniform excitation on the carrier transition $\ket{\uparrow_{l/r}} \leftrightarrow \ket{\downarrow_{l/r}} = \hat{\sigma}^-_{l/r} \ket{\uparrow_{l/r}}$ at frequency $\omega_0$, Rabi-frequency $\Omega_{c}$ and phase $\phi_c$. In the interaction picture and rotating-wave approximation the carrier interaction takes the form
\begin{equation}\label{AEq:CarHam}
\hat{H}_{c} = \hbar \Omega_{c} \left[ (\hat{\sigma}^-_{l}+\hat{\sigma}^-_{r})e^{-i \phi_c}+
(\hat{\sigma}^+_{l}+\hat{\sigma}^+_{r})e^{i \phi_c}\right],
\end{equation}
with $\hat{\sigma}^+_{l/r}=(\hat{\sigma}^-_{l/r})^\dag$. Simultaneously, the ions are driven close to the Raman red sidebands of {\it both} normal modes by two laser beams (index 1 and 2) with difference wave-vector ($\bf k ={\bf k}_2-{\bf k}_1$, magnitude $k= 2 \sqrt{2} \pi/\lambda$) aligned along the direction of the double well, having frequency difference $\Delta \omega_{L} = \omega_2-\omega_1 \approx  \omega_0-\bar{\omega}$, and phase difference $2\phi = k d_0$ of the beat-note between the two laser fields at the positions of the ions. For $\Delta \omega_L = \omega_0$,  the carrier Rabi rate is $\Omega_s$.  We assume the Lamb-Dicke Limit, where  $(\eta_{\rm str/com} q^{(l/r)}_{\rm str/com})^2 \bar{n}_{\rm str/com} \ll 1$, with $\bar{n}_{\rm str/com}$ the average occupation numbers and $\eta_{\rm str/com}=k \sqrt{\hbar/(2 m \omega_{\rm str/com})}$, the Lamb-Dicke parameters of the respective normal modes. The near-resonant terms of the red sideband Hamiltonian are
\begin{eqnarray}\label{AEq:RSBHam}
\hat{H}_{rsb}&=& i \hbar \Omega_{s} \left[
\eta_{\rm com} q_{\rm com}^{(l)} \hat{a}_{\rm com} \hat{\sigma}^+_{l}e^{-i(\delta_{\rm com} t-\phi_s+\phi)}+\eta_{\rm str} q^{(l)}_{\rm str} \hat{a}_{\rm str} \hat{\sigma}^+_{l}e^{-i(\delta_{\rm str} t-\phi_s+\phi)} \right.\nonumber \\
&&\left. + \eta_{\rm com} q^{(r)}_{\rm com} \hat{a}_{\rm com} \hat{\sigma}^+_{r}e^{-i(\delta_{\rm com} t -\phi_s-\phi)}+\eta_{\rm str} q^{(r)}_{\rm str} \hat{a}_{\rm str} \hat{\sigma}^+_{r}e^{-i(\delta_{\rm str} t-\phi_s -\phi)}
\right]+{\rm h.c.},
\end{eqnarray}
where $\delta_{\rm str/com}=\Delta \omega_L-\omega_0+\omega_{\rm str/com}$ is the detuning relative to the red sideband of the respective normal mode, and $\phi_s$ is the phase of the sideband excitation at the mean position of the ions.
\subsection*{Spin-spin interaction}
In the limit of a strongly driven carrier, $|\Omega_{c}| \gg \{ |\eta_{\rm str/com}\sqrt{\bar{n}_{\rm str/com}}\Omega_{s}|, |\delta_{\rm str/com}|\}$, it is helpful to first transform to an internal-state basis where the bare spin states are dressed by the carrier\cite{ber12,tan13}. In this dressed frame, the basis states $\{\ket{+_{l/r}}, \ket{-_{l/r}}\}$ are eigenstates of
\begin{equation}\label{AEq:SigPhi}
\hat{\sigma}_{(l/r)}^{\phi_c}=\cos (\phi_c)\hat{\sigma}^x_{l/r}-\sin (\phi_c) \hat{\sigma}^y_{l/r},
\end{equation}
with $\ket{\pm_{l/r}}=\frac{1}{\sqrt{2}}(\ket{\uparrow_{l/r}} \pm e^{-i \phi_c} \ket{\downarrow_{l/r}})$ and $\hat{\sigma}^{\phi_c}_{l/r}\ket{\pm_{l/r}}=\pm \ket{\pm_{l/r}}$. For each of the four internal basis states $\ket{\pm_{l}} \ket{\pm_{r}}$ and each normal mode, the sideband interaction can be written (neglecting fast oscillating terms near $2\Omega_c$)
\begin{equation}
\hat{H}_{\rm d} = i \hbar (d_{\rm com} e^{i \delta_{\rm com} t} \hat{a}_{\rm com} ^{\dag}- d_{\rm com}^* e^{-i \delta_{\rm com} t} \hat{a}_{\rm com} )+
 i \hbar (d_{\rm str} e^{i \delta_{\rm str} t} \hat{a}_{\rm str}^\dag - d_{\rm str}^* e^{-i \delta_{\rm str} t} \hat{a}_{\rm str}),
\end{equation}
where the coefficients $d_{\rm str/com}$ are state-dependent coherent displacement rates
\begin{equation}\label{AEq:CohDis}
d_{\rm str/com}(s_l,s_r)= -\frac{\Omega_s}{2} \eta_{\rm str/com}(\sin (\theta_{\rm str/com}) s_l e^{-i (\phi_s-\phi_c-\phi)} +\cos (\theta_{\rm str/com}) s_r e^{-i (\phi_s-\phi_c+\phi)}),
\end{equation}
with $s_{l/r} \epsilon \{-1,1\}$ the eigenvalues corresponding to the basis states in question. The integrated displacements $\alpha_{\rm str/com}$ and the geometric phases $\Phi_{\rm str/com}$ acquired after time $t$ are\cite{lei03a}
\begin{eqnarray}\label{AEq:GenDis}
\alpha_{\rm str/com}(s_l,s_r,t)&=& i \frac{d_{\rm str/com}(s_l,s_r)}{\delta_{\rm str/com}}(1-e^{i \delta_{\rm str/com} t}) \nonumber\\
\Phi_{\rm str/com}(s_l,s_r,t)&=&\frac{|d_{\rm str/com}(s_l,s_r)|^2}{\delta_{\rm str/com}^2}(\delta_{\rm str/com} t-\sin (\delta_{\rm str/com} t)).
\end{eqnarray}
To return the motions of both modes to the original state after an interaction duration $T$, we require $\alpha_{\rm str/com}(s_l,s_r,T)=0$. This happens irrespective of the (state-dependent) magnitude of $d_{\rm str/com}$ if $\delta_{\rm str/com} T = c_{\rm str/com} (2 \pi)$ with $c_{\rm str/com}$ integer.  In such cases the motion is displaced around $|c_{\rm str/com}|$ full circles in the respective phase spaces of both modes by the interaction. Also, since $\delta_{\rm str}-\delta_{\rm com}=2 \sqrt{\delta^2+\Omega_{\rm ex}^2}$, the interaction duration can assume only certain values determined by $\Delta c \equiv c_{\rm str} - c_{\rm com} >0$ for the motion to return to its original state:
\begin{equation}\label{AEq:GatTim}
T=\frac{\pi \Delta c}{\sqrt{\delta^2+\Omega_{\rm ex}^2}}.
\end{equation}
If the spin and motional states are in a product state initially, they will be in a product state at $T$ and any integer multiple of $T$. The  spin-dependent phases acquired during $T$ simplify to 
\begin{equation}\label{AEq:LooPha}
\Phi_{\rm str/com}(s_l,s_r)=\left(\frac{\eta_{\rm str/com} \Omega_s}{2} \right)^2 \frac{T}{\delta_{\rm str/com}}[1+ s_l s_r \cos( 2\phi)\sin(2 \theta_{\rm str/com})].
\end{equation}
The spin-dependent term is largest if $\phi= j \pi/2$ with $j$ integer. This corresponds to the ions being spaced by an integer number of half-wavelengths $ \pi/k$. In the experiment, the separation of the ions is controlled by slight changes in the well curvatures to ensure half-integer wavelength spacing. Also, $|\sin(2 \theta_{\rm str/com})|$ is reduced for $|\delta|>0$ and eventually vanishes as the modes decouple in the limit $|\delta| \gg \Omega_{\rm ex}$; therefore, the most efficient spin-spin interactions are implemented for $\delta=0$.  For our experimental conditions and $\delta=0$, the mode splitting is much smaller than the average mode frequency $\bar{\omega}$, so we can approximate $\eta_{\rm str/com} \approx \eta =k \sqrt{\hbar/(2 m \bar{\omega})}$. If we also assume that $\delta \ll \Omega_{ex} $, the phases simplify to 
\begin{equation}\label{AEq:PhaApp}
\Phi_{\rm str/com}(s_l,s_r,T)=\left(\frac{\eta \Omega_s}{2} \right)^2 \frac{T}{\delta_{\rm str/com}}\left[1\mp s_l s_r \cos( 2\phi)\left(1-\frac{\delta^2}{2 \Omega_{\rm ex}^2}\right)\right].
\end{equation}
In this limit, the phases $\Phi_{\rm str/com}(s_l,s_r,T)$ depend only to second order on the relative detuning of the two wells. The shortest loop duration $T$ is realized for $\Delta c =1$, but the phase accumulates most effectively when the sideband drive is tuned to $\bar{\omega}$, exactly halfway between the normal modes ($c_{\rm str/com}\!=\!\pm 1, \Delta c\!=\!2$). At this detuning the logical phase acquired on both modes adds constructively, while there is always some degree of phase cancellation for all other possible settings of the detuning. The total phase accumulated on both modes during $T$ is
\begin{equation}\label{AEq:TotPha}
\Phi(s_l,s_r,T)=\Phi_{\rm str}(s_l,s_r,T)+\Phi_{\rm com}(s_l,s_r,T) = - \cos( 2\phi)\frac{(\eta \Omega_s)^2}{2 \Omega_{\rm ex}} s_l s_r  T .
\end{equation}
For any integer multiple of $T$, we can summarize the action of the applied fields as
\begin{equation}\label{AEq:ActSta}
\ket{\pm_l, \pm_r, jT} = \exp[- i \cos( 2\phi)\frac{(\eta \Omega_s)^2}{2 \Omega_{\rm ex}} \hat{\sigma}^{\phi_c}_{l} \hat{\sigma}^{\phi_c}_{r} jT] \ket{\pm_l, \pm_r, 0},
\end{equation}
with $j$ a positive integer. Since this holds for a complete set of spin-basis states, it also holds for any general initial state of the system. Therefore, at any multiple of $T$, the system evolution is equivalent to that under the spin-spin Hamiltonian
\begin{eqnarray}\label{AEq:SpiHam}
\hat{H}_{\rm eff}&=& \hbar \kappa \hat{\sigma}^{\phi_c}_{l} \hat{\sigma}^{\phi_c}_{r}\nonumber\\
\kappa&=&\cos( 2\phi)\frac{(\eta \Omega_s)^2}{2 \Omega_{\rm ex}}.
\end{eqnarray}
A change from ferromagnetic to anti-ferromagnetic interaction can be accomplished by a $\pi/k$ change of the ion spacing, corresponding to a $\pi/2$ change of $\phi$. Alternatively, for example, $\kappa' = -\kappa/3<0$ is realized with a choice of detuning such that ($c_{\rm str}=-1, c_{\rm com}=-3$).

In principle, we can either perform a `stroboscopic' emulation with the total duration a multiple of $T$, or use detunings $\delta_{\rm str/com}$ whose magnitudes are much larger, so that all $|\alpha_\pm| \ll 1$ for any given time. For all multiples of $T$, the motional states of the ions factor from the spin states, so if one only ``looks'' stroboscopically at times $j T$, the system effectively appears as though only the spins have evolved according to Eq.(\ref{AEq:SpiHam}), while the motion has returned to its original state, thus appearing to have been unaffected.  For much larger magnitude detunings $\delta_{\rm str/com}$ spin-motion entanglement, and thus the deviation of the simulated state from that under the ideal spin-spin interaction, is small for arbitrary durations of the interaction\cite{por04}. The added robustness comes at the expense of a weaker spin-spin interaction that has to be compensated by higher drive power or longer simulation time scales. Finally, rather than suppressing the bosonic harmonic oscillator modes, we can include them as an integral part of the simulator and study collective spin-boson Hamiltonians, which have been recently shown to contain complex behavior comparable to models with only spin-spin interactions \cite{jun13}.
\subsection*{Experimental characterization}
We benchmark the spin-spin Hamiltonian of Eq.(\ref{AEq:SpiHam}) by using it to entangle the hyperfine states (pseudo-spins) of two ions starting from the initial state $\ket{\downarrow \downarrow}$. To gain isolation from small errors, we break the total spin-spin interaction into two loops in phase space with $\kappa=\pi/8$ for each loop. For the first loop we can choose $\phi_c=0$ and $\phi_s=0$ so that the eigenstates in the dressed basis are those of $\hat{\sigma}^x_{l/r}$. After finishing the first loop, we change carrier and sideband phase to $\phi_c=\phi_s=\pi$. The change in carrier phase realigns the rotating frame due to the carrier with the frame of the bare spin states at the end of the second loop as the rotations around the $x$-axis of the Bloch sphere in the first loop are unwound by rotating around the $-x$-axis for the same duration in the second loop. In addition, the phase change in the sideband drive ensures that $d_{\rm str/com}(s_l,s_r)$ of the first loop is followed by $-d_{\rm str/com}(s_l,s_r)$ in the second loop. In total there are three sign changes in the displacement rate Eq.(\ref{AEq:CohDis}), the first from replacing $\hat{\sigma}^x_{l/r}$ by $\hat{\sigma}^{-x}_{l/r}=-\hat{\sigma}^x_{l/r}$ and therefore $s_{l,r}\rightarrow -s_{l,r}$, the second due to $\phi_c=0 \rightarrow \pi$ and the third due to $\phi_s=0 \rightarrow \pi$, which multiply to change the sign of the displacement rate. As a consequence, the total displacement $\alpha_{\rm str/com}(s_l,s_r,T)$ in the second loop [see Eq.(\ref{AEq:GenDis})] is equal and opposite to that in the first loop and the motional wave functions return to their original position in phase space even if $\alpha_{\rm str/com}(s_l,s_r,T) \neq 0$ due to small errors in detunings $\delta_{\rm str/com}$ or loop duration, providing those errors are constant over both loops \cite{hay12}. The phases $\Phi_{\rm str/com}(s_l,s_r)$ depend only on $|d_{\rm str/com}(s_l,s_r)|^2$, therefore the effective spin-spin evolution is the same in both loops. With the sideband excitation tuned to $\bar{\omega}$, a single loop duration corresponds to $T_L=2 \pi/\Omega_{\rm ex}$ for a total interaction duration of $2 T_L$. Starting from  the initial state $\ket{\downarrow \downarrow}$  we would ideally produce the maximally entangled state $\ket{\Psi_e}=\exp[-i \frac{\pi}{4}\hat{\sigma}^x_{l}\hat{\sigma}^x_{r} ]\ket{\Psi_i}=\frac{1}{\sqrt{2}} (i \ket{\downarrow \downarrow} - \ket{\uparrow \uparrow})$, if the sideband Rabi-frequency fulfills $\eta \Omega_s =\Omega_{\rm ex}/(2 \sqrt{2})$.
\subsection*{Determination of probabilities from state-dependent fluorescence}
During one detection period (duration 300 to 400 $\mu$s) we typically detect between 0.15 and 0.6 counts if both ions are projected into $\ket{ \uparrow}$, and 3 to 5 additional counts for each ion in state $\ket{ \downarrow}$. For each experimental setting, we record count histograms for 200 to 500 experiments.

Consider a count histogram $h=(h(i))_i$, where $h(i)$ experiments yielded $i$ counts and $N=\sum_i h(i)$ is the total number of recorded counts.  We infer the probabilities $P_b$ with $b=0,1,2$ by applying probability estimators $w_b=(w_b(i))_i$ to $h$ according to $P_b = \sum_i w_b(i)h(i)/N$. The probability estimators are determined from the recorded photon counts for on-resonance microwave Ramsey experiments with two ions, where the phase $\phi$ of the second $\pi/2$-pulse was
varied. These experiments are performed before and after the experiments to be analyzed. An ideal such Ramsey experiment satisfies
\begin{eqnarray}
P_0(\phi)=\cos^4(\phi/2) \nonumber \\
P_1(\phi)=\sin^2(\phi)/2 \nonumber \\
P_2(\phi)=\sin^4(\phi/2).
\end{eqnarray}
The histograms $h_\phi$ recorded at phase $\phi$ are sampled from the mixture $P_0 q_0+P_1 q_1+P_2 q_2$, where the $q_b$ are the count distributions for zero, one or two ions bright.  From this model and the Ramsey data, we can determine $w_b$ so that $\sum_i w_b(i) h_\phi(i)$ yields $P_b(\phi)$.  We use a linear-least-squares method fit, regularizing  it to minimize the anticipated variance when inferring $P_b$ for the completely mixed state.

Given a probability estimator $w$ and a recorded histogram $h$, we estimate the experimental variance of the inferred probability $P$ according to $v = (\sum_i w(i)^2 h(i)/N - P^2)/(N-1)$.  This variance determines the error bars in Fig.~3.  For the fidelities and related quantities, the variation in the probability estimators due to finite statistics of the Ramsey experiments contributes an error comparable to this variance. To determine the overall statistical error in the fidelities, we used non-parametric bootstrap resampling~\cite{efron} on all contributing histograms with $100$ bootstrap resamples to determine error bars for fidelities and contrasts.

The assumed model for the Ramsey experiments makes no assumptions on the shapes or relationships of the count distributions $q_b$.  This was important because we found that the $q_b$ exhibit clear deviations from Poissonian distributions.  We also determined $c_b$, the mean number of counts according to $q_b$, and found that $c_2-c_0$ exceeded $2 \times (c_1-c_0)$ by about $8\;\%$ for all the Ramsey scans considered.

Several effects result in deviations from an ideal Ramsey experiment. We found that there is a phase offset of approximately $5^{\circ}$ in the Ramsey scans. We shifted the phase accordingly before determining the
probability estimators. This had a statistically negligible effect on inferred probabilities and fidelities.  After adjusting for the phase shift, we found no signature of a mismatch between the model and the data. In addition to checking that the dependence of the histograms on the phase was as expected, we considered whether there are more than three count distributions contributing to the Ramsey scans. We found no signature of such an effect. Furthermore, all other histograms, including those used to determine fidelities, could be explained as arising from a mixture of the same three count distributions.

An important effect that need not be apparent from the data is state-preparation error. By simulating Ramsey experiments with state-preparation error and $q_b$ as inferred from the data, we determined
that such errors lead to systematic overestimates of fidelities that are
well-correlated with the state preparation error.  The simulations
involved initial states that are mixtures of the basis states.  Let
$\epsilon~(\ll1)$ be the probability that the state in this mixture is not
$\ket{\downarrow\downarrow}$. For the inferred fidelities, we estimate
a systematic increase in fidelity of approximately $1.1\times\epsilon$. The quoted systematic
errors are based on a pessimistic upper bound of $0.01$ on $\epsilon$. For
inferring the $P_b$ for a single histogram (as required for the plots
in Fig.~3), these biases are small compared to the
statistical error and were therefore not included in the error bars.
We assumed that pulse errors had a statistically small effect on
inferred probabilities and fidelities.

\end{document}